\documentclass{raa}            
\usepackage{graphicx,times}    
\usepackage{amssymb,amsmath,amsfonts}
\usepackage{makecell,multirow,textcomp}
\usepackage{float,placeins}
\usepackage{enumerate,enumitem}
\usepackage{multicol,mathrsfs}
\usepackage{etoolbox}
\usepackage{tabularx}
\usepackage{url}
\usepackage[english]{babel}
\usepackage[nottoc]{tocbibind}
\usepackage{authblk}
\pdfoutput=1
 
\renewcommand{\arraystretch}{1.5}
 
\graphicspath{{figures/}} 
\usepackage[shortcuts]{extdash}
\usepackage{epstopdf} 
\usepackage{natbib}
\bibpunct{(}{)}{;}{a}{}{,}
 
\usepackage{pdfpages}

\makeatletter
\patchcmd{\section}{\clearpage}{}{}{} 
\patchcmd{\subsection}{\clearpage}{}{}{} 
\makeatother
\usepackage{color}
\usepackage[pagebackref=true]{hyperref}
\hypersetup{ colorlinks=true, linkcolor=blue, filecolor=magenta,       urlcolor=blue,citecolor=blue, pdftitle={Your Paper Title}}

\begin{document}
   \title{Inferences for $f(R)$ Models from Late-time Megamaser Observational Data}

   \volnopage{ {\bf 20XX} Vol.\ {\bf X} No. {\bf XX}, 000--000}
   \setcounter{page}{1}
      
   \author{Umang\inst{1}, Abha Dev Habib\inst{1} \and Nisha Rani\inst{1} }

\authorrunning{Umang et al.} 
\titlerunning{Inferences for $f(R)$ Models}
\institute{
    $^1$ Department of Physics, Miranda House, University of Delhi, University Enclave, Delhi-110007, India;
    {\it abhadev.habib@mirandahouse.ac.in}\\
    \vs \no
    {\small Received 20xx month day; accepted 20xx month day}
}

 \abstract{ In this work, we study three widely used models of $f(R)$ gravity, namely Hu-Sawicki, Starobinsky and ArcTanh along with the standard cosmology model ($\Lambda$CDM). For this, we employ the  megamaser angular diameter distance and velocity measurements from the Megamaser Cosmology Project, which provide a purely geometric determination of the Hubble constant. We constrain the parameters using the Markov Chain Monte Carlo method. Our results show that values of the Hubble Constant, $H_0$, obtained for all four models are in concordance with  its value obtained from other late-time observational data such as SNe Ia. The constraints on $H_0$ in all the models under study are restrictive and the marginalized estimates lie close to 73\,km s$^{-1}$\,Mpc$^{-1}$. The marginalized estimates of the deviation parameter, $b$, for the three $f(R)$ gravity models lie close to zero. This late-time dataset, thus predict that $f(R)$ models mimic $\Lambda$CDM. However, the matter density, $\Omega_{m}$, remains weakly constrained for all the models with its marginalized estimate close to 0.5. Further, comparison of the four models ($f(R)$ models and $\Lambda$CDM) using information criteria such as Akaike Information Criterion and Bayesian Information Criterion shows that within current uncertainties, the dataset finds $f(R)$ models statistically indistinguishable from $\Lambda$CDM. This is consistent with the fact that the favoured value of $b$ for each of the $f(R)$ models  lies close to zero.
}

\maketitle 

\section{Introduction}
\label{introduction}
The spatially flat $\Lambda$CDM model,  the standard model of cosmology, incorporates baryonic matter, cold dark matter and a cosmological constant (a form of dark energy). Though it provides an excellent description of a wide range of cosmological observations, many  tensions remain unresolved. To list a few, the Hubble constant discrepancy, indications of inconsistencies in the spatial curvature parameter, and theoretical concerns such as the fine-tuning and coincidence problems are major unsolved mysteries (\citealt{2000astro.ph..5265W,2003RvMP...75..559P,2021ApJ...919...16F,2023OJAp....6E..36D}). These issues provide strong motivation to go beyond the standard model.\\
\indent For this reason, a wide variety of theoretical efforts aim at addressing the observed discrepancies. Some proposals introduce non-standard dark energy components by making the dark energy component dynamic. This includes quintessence models in which dark energy is driven by a scalar field $(\phi)$ and is considered to be slowly rolling down a potential energy landscape, k-essence fields which are driven by non-standard kinetic terms in the Lagrangian and phantom dark energy models which allow the equation of state to be less than $-1$ (\citealt{2013CQGra..30u4003T,2017MPLA...3230025L,2023ForPh..7100133M}). Though these non-standard models were introduced to address the theoretical issue of $\Lambda$ but they introduce some new and more severe problems namely the fine-tuning problem, the vacuum instability problem and causality violation (\citealt{2001PhRvD..63j3510A,2002PhLB..545...23C,2003PhRvD..68b3509C,amendola2010dark,2012Ap&SS.342..155B}).\\
\indent An alternate and more direct strategy is to focus on modifying the dark energy equation of state through phenomenological parametrizations. In this approach, the dark energy equation of state is parametrized in various forms like the Chevallier–Polarski–Linder,  Jassal-Bagla-Padmanabhan, Feng-Shen-Li-Li, and Barboza-Alcaniz parametrization (\citealt{2001IJMPD..10..213C,2003PhRvL..90i1301L,2005MNRAS.356L..11J,2008PhLB..666..415B,2012JCAP...09..023F}). The problems with this approach are that such extensions generally introduce additional parameter degeneracies that can weaken observational constraints. A more fundamental approach is provided by inhomogeneous cosmological constructions, exemplified by the $\text{Lema$\hat{\text{i}}$tre}$\-/Tolman\-/Bondi solution (\citealt{pascual1999cosmic}). This suggests that large-scale spatial inhomogeneities may effectively reproduce signatures commonly attributed to dark energy without invoking exotic components. Nevertheless, these scenarios typically rely on finely tuned observer configurations and face increasingly stringent bounds from current observational data (\citealt{2016arXiv160105256S}).\\
\indent Beyond phenomenological dark energy parametrizations and inhomogeneous cosmological scenarios, modified gravity frameworks offer another alternative which is geometrically motivated. This is done by extending the Einstein–Hilbert action through higher-order curvature corrections. In its simplest realization, General Relativity (GR) is generalized by replacing the Ricci scalar (R) with an arbitrary function $f(R)$ (\citealt{2010LRR....13....3D,2012PhR...513....1C}). This construction was initially explored in the context of inflation and is now being extensively investigated as a mechanism for late-time acceleration of the universe. Within this class of theories, accelerated expansion arises from the underlying spacetime dynamics rather than from an explicit dark energy sector. Here, it is important to construct viable models carefully in order to remain consistent with both cosmological observations and local gravity constraints (\citealt{2003PhRvD..68l3512N,2007IJGMM..04..115N,2011PhR...505...59N,2007PhRvD..75h3504A,2008PhRvD..78h3515D,2010RvMP...82..451S,2017PhR...692....1N}). Keeping this in mind, a wide range of functional forms have been proposed, often designed to reproduce $\Lambda$CDM behavior in appropriate limits while retaining distinctive cosmological signatures.\\
\indent For any theoretical model to be accepted, it is important that it should be consistent with observations. For that reason, current research in $f(R)$ gravity is increasingly driven by observational viability. The primary goal is to identify observationally consistent models that (i) reproduce the well-tested phenomenology of $\Lambda$CDM, (ii) satisfy solar-system bounds, and (iii) generate measurable deviations in the growth of large-scale structure and cosmic expansion history (\citealt{2001PhLB..506...13H,2007PhRvD..75d4004S}). The difficulty while constructing a theoretical functional form is that it should also be consistent with high-precision cosmological probes like supernovae Ia, baryon acoustic oscillations, and cosmic microwave background. While early curvature-corrected $f(R)$ models successfully captured aspects of inflationary dynamics, and inverse of the curvature proposals for late-time acceleration were largely excluded by stability and local-gravity constraints, modern viable formulations are explicitly designed to pass both background and perturbative observational tests (\citealt{STAROBINSKY198099,2003PhLB..575....1C,2003PhLB..573....1D,2003PhRvD..68l3512N,2004PhRvD..70d3528C}). These developments bring out the potential of $f(R)$ models to provide a unified description of cosmic acceleration while remaining consistent with the full spectrum of current observational data.\\
\indent Observational constraints on $f(R)$ gravity models have been extensively explored using a wide range of complementary cosmological observations. Early studies concentrated on the Hu–Sawicki framework, combining measurements of X-ray cluster gas mass fractions with cosmic microwave background and BAO distance ratios, together with the Union 2.1 supernova compilation, to place joint limits on model parameters (\citealt{2011PhLB..699..320C,2012A&A...548A..31S}). The field has progressed along comprehensive multi-model analyses which includes growth-rate data, Type Ia supernova (SNe Ia) observations and Hubble constant data (\citealt{2013PhRvD..87l3529B,2017JCAP...01..005N,2018PhRvD..97b3525P,2018arXiv180702163O}). More recent studies have significantly strengthened the models by incorporating additional observations namely gravitational-lensing, quasar data, $H_{\text{II}}$ galaxy measurements, non-parametric Gaussian-process reconstructions of the Hubble diagram along with the joint likelihood analyses (\citealt{2019PhRvD.100d4041D,2020PhRvD.101j3505D,2021NuPhB.96615377O,2022PhRvD.105j3526L,2022MNRAS.514.5827S,2025ApJ...992..205B,2024arXiv240504886R}). Recently, \cite{kumar2025joint} constrained $f(R)$ models using joint analysis of both early and late-time cosmological datasets. This framework offers a compelling phenomenological alternative to conventional dark energy explanations of cosmic acceleration. The observed late-time accelerated expansion is attributed to modifications in spacetime dynamics itself instead of introducing additional exotic energy components.\\
\indent To take the study of $f(R)$ models further, we employ here the megamasers data collected by the Megamaser Cosmology Project (MCP), also referred to as the P20 project (\citealt{2020ApJ...891L...1P}). The MCP is meant to map, collect and analyze megamaser data to constrain the Hubble parameter ($H_{0}$). Megamasers are extremely luminous masers that are used to analyze the far off galaxies.
Water megamasers, which are the most studied, are found in the accretion disk around supermassive black holes (SMBHs) in the Active Galctic Nucleus (AGN) region of galaxy. Water megamasers provide a way to measure the Hubble constant $H_{0}$ in the low redshift range. The inferred value of $H_{0}$ is independent of the distance ladder measurements and cosmic microwave background measurements (\citealt{2020ApJ...891L...1P,2025Ap&SS.370...62B}). For the  same reason, it is important to investigate cosmological models using the megamaser data. \\
\indent This paper is organised as follows: We discuss the data and methodology in  Section~\ref{Data and Methodology}. The P20 dataset, Bayesian Analysis, $f(R)$ models and Markov Chain Monte Carlo Method are discussed in Section~\ref{The P20 Dataset}, \ref{Bayesian Analysis}, \ref{f(R) models:} and \ref{MCMC method} respectively. In Section~\ref{results}, we present our results obtained from the analysis. We finally conclude our findings in Section~\ref{conclusion}.

\FloatBarrier
\section{Data and Methodology} 
\label{Data and Methodology}
\subsection{The P20 Dataset}
\label{The P20 Dataset}
The P20 dataset is emerging as an important probe to constrain cosmological parameters as it provides a purely geometric determination of the Hubble constant (\citealt{1999Natur.400..539H}). It uses  Very Long Baseline Interferometry (VLBI) observations of water megamasers hosted in nearly edge-on, Keplerian accretion disks around SMBHs, enabling direct measurements of angular-diameter distances (\citealt{2009ApJ...695..287R,2019ApJ...886L..27R}). The P20 analysis utilizes six well-modelled megamaser galaxies (UGC 3789, NGC 6264, NGC 6323, NGC 5765b, CGCG 074-064 and NGC  4258), spanning both local and Hubble-flow regimes, with appropriate corrections for peculiar velocities. Owing to its geometric nature and minimal astrophysical systematics, determination from the P20 dataset serves as an important late-time anchor for cosmological analysis.   \\
\indent The dataset used for this analysis includes  data points corresponding to the six megamaser host galaxies. The dataset comprises of their angular diameter distances, $\hat{D_{A}}$, their recession velocities, $\hat{v_i}$ and the associated uncertainties in $\hat{D_{A}}$ and $\hat{v_i}$ (see Table \ref{megamaserdata}). In the dataset, taken from \cite{2020ApJ...891L...1P}, megamaser galaxies lie in the redshift range of
0.002$<$z$<$0.05. Here, $z$ refers to the cosmological redshift. For low redshift (z$<<$ 1), it is related to the expected cosmological recession velocity by the equation:
\begin{equation}
    z_{i}=\frac{\text{}v_i}{c},
    \label{redshift_veocity}
\end{equation}
\noindent where $c$ is the speed of light in vacuum. \cite{2020ApJ...891L...1P} mentioned that the expected cosmological recessional velocity differs from the measured galaxy velocity. It is because of the statistical uncertainty in the measurement and systematic uncertainty contributed due to peculiar motion. While the statistical uncertainties in the measurement are very small, i.e., of the order of 1-2\,km s$^{-1}$, the recession velocity is dominated by the peculiar motion as the galaxies reside at very low redshift. In order to take care of the peculiar velocities, we follow the approach of \cite{2020ApJ...891L...1P} and incorporate them into the velocity measurement uncertainty  (see Equation(\ref{chi2})).
\begin{table}[H]
\begin{center}
\caption{P20 dataset Comprised of Angular Diameter Distance, $\hat{D_{A}}$ and Recession Velocity, $\hat{v_i}$ for the Megamaser Host Galaxies.}
\label{megamaserdata}
\renewcommand{\arraystretch}{1} 
 \begin{tabular}{clclclcl}
 \hline\noalign{\smallskip}
        No.&Galaxy & $\hat{D_{A}}$ (Mpc)& $\hat{v_i}$ (km s$^{-1}$)\\
         \hline\noalign{\smallskip}
 		1&UGC 3789 &  51.5 $\pm$ 4.2 & 3319.9 $\pm$ 0.8 \\
 		2&NGC 6264 & 132.1 $\pm$ 19 & 10192.6 $\pm$ 0.8 \\
 		3&NGC 6323 & 109.4 $\pm$ 28.5 & 7801.5 $\pm$ 1.5 \\
 		4&NGC 5765b & 112.2 $\pm$ 5.2& 8525.7 $\pm$ 0.7 \\
 		5&CGCG 074-064& 87.6 $\pm$ 7.5 & 7172.2 $\pm$ 1.9 \\
 		6&NGC 4258 & 7.58 $\pm$ 0.11 & 679.3 $\pm$ 0.4 \\ 
        \hline\noalign{\smallskip}
\end{tabular}
\end{center}
\end{table}
\FloatBarrier
\subsection{Bayesian Analysis} 
\label{Bayesian Analysis}
We use combined likelihood L for our analysis 
\begin{equation}
    \rm \textit L= \textit L_{\textit{v}}\,\textit L_{\textit{D}}. 
\end{equation}
 Here, $\textit L$$_{v}$ is the velocity contribution to the likelihood and is to account for the systematic uncertainty owing to peculiar motion. $\textit L$$_{D}$ is the contribution to the likelihood from the distance measurements.\\
\indent The likelihood function is defined as 
\begin{equation}
    \text{}
    \rm ln \,\textit L=-0.5\chi^{2},
    \label{likelihood}
\end{equation}
where
 \begin{equation}
     \quad \chi^{2}=\sum_{i}\left[\frac{(v_{i}-\text{}\hat{v_i})^{2}}{\sigma^{2}_{\rm pec}+\sigma^{2}_{\text{}\hat{v_i}}}+\frac{(D_{\rm A}-\text{}\hat{D_A})^{2}}{\sigma^{2}_{\text{}\hat{D_A}}}\right].
 \label{chi2}
 \end{equation}
  Here, $\hat{D_{A}}$ and $\sigma_{\hat{D_A}}$ are the observed angular diameter distance and their related uncertainty, respectively. $D_{\rm A}\text{}$ represents the theoretical value of the angular diameter distance. $\hat{v_i}$  and $v_{i}\text{}$  represent the observed and theoretical recession velocities of $i$th host galaxy respectively. The uncertainty associated with the observed $\hat{v_i}$ is represented by $\sigma_{\hat{v_i}}$. $\sigma_{\rm pec}$ represents the uncertainty owing to the peculiar velocity of the galaxy.  Given that peculiar velocities of galaxies lie in the range 150-250\,km s$^{-1}$, following \cite{2020ApJ...891L...1P}, $\sigma_{\rm pec}$ is taken to be 250\,km s$^{-1}$. Fixing the upper end of the range as a common peculiar velocity uncertainty for the entire sample may be a conservative assumption. However this is done to ensure that the constraints on parameters are within the bounds of maximum possible uncertainties. Apart from the model parameters, Equation (\ref{chi2}) is used to constrain $v_i\text{}$.\\
\indent The theoretical angular diameter distance is:
  \begin{equation}
  D_{\rm A}\text{}(z\text{$, p$})=\frac{c}{(1+z)}\int_{0}^{z}\frac{dz'}{H(z'\text{, p})}.
  \label{dA}
 \end{equation}
 Here, $H(z, p)$ is the Hubble parameter which apart from the model parameters depends on $z$. The low redshift approximation given by Equation \eqref{redshift_veocity} is used to convert $D_A$ as a function of $v_i$ and model parameters $p$.\\
\indent For $\Lambda$CDM, with $\textit p$ as $\left\{H_0, \Omega_m \right\}$,  the Hubble parameter $H(z\text{$, p$})$ is given as:
\begin{equation}
\begin{aligned}
H(z\text{$, p$})&=H_{0}\sqrt{\Omega_{\rm m}(1+z)^{3}+(1-\Omega_{\rm m})}.\\
\end{aligned}
\label{CDM model eq}
\end{equation}

\subsection{\textit{f(R)} Models}
\label{f(R) models:}
\noindent For the $f(R)$ models considered in this work, i.e., Hu-Sawicki, Starobinsky and ArcTanh, we use the expression of $H(z\text{$, p$})$ that relates to $z$ and model parameters , $p\,=\, \left\{H_{0}, b, \Omega_{\rm m}\right\}$. (for details, see \cite{2022MNRAS.514.5827S}).

\subsubsection{Hu-Sawicki Model}
It is a widely studied viable extension of GR within the framework of $f(R)$ modified gravity. The model was originally proposed to explain late-time cosmic acceleration without invoking a cosmological constant. In this model, the Einstein–Hilbert action is generalized by replacing the Ricci scalar (R) with a nonlinear function $f(R)$ (\citealt{2007PhRvD..76f4004H}).
For this model, the Hubble parameter $H(z\text{$, p$})$ is
\begin{equation}
\begin{aligned}
   \frac{H(z\text{$, p$})^{2}}{H_{0}^{2}}&=1-\Omega_{\rm m}+(1+z)^{3}\Omega_{\rm m}\\&+{6b(\Omega_{\rm m}-1)^{2}}\left[\frac{4(\Omega_{\rm m}-1)^{2}+(1+z)^{3}(\Omega_{\rm m}-1)\Omega_{\rm m}-2(1+z)^{6}\Omega_{\rm m}^{2}}{(1+z)^{9}\left(\frac{4(\Omega_{\rm m}-1)}{(1+z)^{3}}-\Omega_{\rm m}\right)^{3}}\right]\\
    &+ \frac{b^{2}(\Omega_{\rm m}-1)^{3}}{(1+z)^{24}\left(\frac{4(1-\Omega_{\rm m})}{(1+z)^{3}}+\Omega_{\rm m}\right)^{8}}\biggl[1024(\Omega_{\rm m}-1)^{6}+9216(1+z)^{3}(\Omega_{\rm m}-1)^{5}\Omega_{\rm m}\\&-22848(1+z)^{6}(\Omega_{\rm m}-1)^{4}\Omega_{\rm m}^{2}
  +25408(1+z)^{9}(\Omega_{\rm m}-1)^{3}\Omega_{\rm m}^{3}\\&- 7452(1+z)^{12}(\Omega_{\rm m}-1)^{2}\Omega_{\rm m}^{4}-4656(1+z)^{15}(\Omega_{\rm m}-1)^{1}\Omega_{\rm m}^{5}+37(1+z)^{18}\Omega_{\rm m}^{6}\biggr].
\end{aligned}
\label{Hu-Sawicki model eq}
\end{equation}
\subsubsection{Starobinsky Model}
It is a well-known $f(R)$ modified gravity model. This model extends GR by including higher-order curvature corrections to offer a geometric explanation for cosmic acceleration. The model incorporates nonlinear curvature terms that drive the accelerated expansion phase without an explicit scalar field requirement. The Starobinsky-type $f(R)$ function is designed to closely mimic the $\Lambda$CDM expansion history while simultaneously satisfying the local gravity constraints (\citealt{2007JETPL..86..157S}).
For Starobinsky model, the Hubble parameter $H(z\text{$, p$})$ is
\begin{equation}
\begin{aligned}
   \frac{H(z\text{$, p$})^{2}}{H_{0}^{2}}&=1-\Omega_{\rm m}+(1+z)^{3}\Omega_{\rm m}\\&+b^{2}(\Omega_{\rm m}-1)^{3}\left[\frac{32(\Omega_{\rm m}-1)^{2}+32(1+z)^{3}(\Omega_{\rm m}-1)\Omega_{\rm m}-37(1+z)^{6}\Omega_{\rm m}^{2}}{(1+z)^{12}\left(-\frac{4(\Omega_{\rm m}-1)}{(1+z)^{3}}+\Omega_{\rm m}\right)^{4}}\right]\\
    &+ \frac{b^{4}(\Omega_{\rm m}-1)^{5}}{(1+z)^{30}\left(\frac{4(1-\Omega_{\rm m})}{(1+z)^{3}}+\Omega_{\rm m}\right)^{10}}\biggl[20480(\Omega_{\rm m}-1)^{6}+63488(1+z)^{3}(\Omega_{\rm m}-1)^{5}\Omega_{\rm m} \\&-234880(1+z)^{6}(\Omega_{\rm m}-1)^{4}\Omega_{\rm m}^{2}
  +289024(1+z)^{9}(\Omega_{\rm m}-1)^{3}\Omega_{\rm m}^{3}\\&- 44552(1+z)^{12}(\Omega_{\rm m}-1)^{2}\Omega_{\rm m}^{4}-82748(1+z)^{15}(\Omega_{\rm m}-1)^{1}\Omega_{\rm m}^{5}+123(1+z)^{18}\Omega_{\rm m}^{6}\biggr].
\end{aligned}
\label{Starobinsky model eq}
\end{equation}
\subsubsection{ArcTanh Model}
The ArcTanh model of $f(R)$ gravity is a phenomenological modification of GR. In this model, the Ricci scalar, R, is replaced by a function that contains an ArcTanh dependence. It is designed to produce late-time cosmic acceleration while it is consistent with local gravity tests. This form allows the modification to interpolate smoothly between low and high curvature regimes. At cosmological scales, this model introduces the behavior of dark-energy without explicitly invoking a cosmological constant.
Owing to its well-behaved functional form, the ArcTanh model provides a useful framework to explore stable deviations from standard gravity and testing modified gravity models against observational data (\citealt{2018PhRvD..97b3525P}). The Hubble parameter $H(z\text{$, p$})$  is given as

\begin{equation}
\begin{aligned}
   \frac{H(z\text{$, p$})^{2}}{H_{0}^{2}}&=1-\Omega_{\rm m}+(1+z)^{3}\Omega_{\rm m}+\frac{2b(\Omega_{\rm m}-1)^{2}}{3(1+z)^{15}\left(\frac{4(\Omega_{\rm m}-1)}{(1+z)^{3}}-\Omega_{\rm m}\right)^{5}} \biggl[596(\Omega_{\rm m}-1)^{4}\\&-109(1+z)^{3}(\Omega_{\rm m}-1)^{3}\Omega_{\rm m}-361(1+z)^{6}(\Omega_{\rm m}-1)^{2}\Omega_{\rm m}^{2}+153(1+z)^{9}(\Omega_{\rm m}-1)\Omega_{\rm m}^{3}\\&-18(1+z)^{12}\Omega_{\rm m}^{4} \biggr]+ \frac{b^{2}(\Omega_{\rm m}-1)^{3}}{9(1+z)^{36}\left(\frac{4(1-\Omega_{\rm m})}{(1+z)^{3}}+\Omega_{\rm m}\right)^{12}}\biggl[4189184(\Omega_{\rm m}-1)^{10}\\&+23851008(1+z)^{3}(\Omega_{\rm m}-1)^{9}\Omega_{\rm m} -95032704(1+z)^{6}(\Omega_{\rm m}-1)^{8}\Omega_{\rm m}^{2}
  \\&+149808704(1+z)^{9}(\Omega_{\rm m}-1)^{7}\Omega_{\rm m}^{3}-110911572(1+z)^{12}(\Omega_{\rm m}-1)^{6}\Omega_{\rm m}^{4}\\&+28668900(1+z)^{15}(\Omega_{\rm m}-1)^{5}\Omega_{\rm m}^{5}+2987537(1+z)^{18}(\Omega_{\rm m}-1)^{4}\Omega_{\rm m}^{6}\\&-3144240(1+z)^{21}(\Omega_{\rm m}-1)^{3}\Omega_{\rm m}^{7}+636102(1+z)^{24}(\Omega_{\rm m}-1)^{2}\Omega_{\rm m}^{8}\\&-47232(1+z)^{27}(\Omega_{\rm m}-1)\Omega_{\rm m}^{9}+333(1+z)^{30}\Omega_{\rm m}^{10}\biggr].
\end{aligned}
\label{ArcTanh model eq}
\end{equation}
\subsection{Markov Chain Monte Carlo (MCMC) Method}
\label{MCMC method}
In this study, we employ MCMC method based on Bayesian statistics to put constraints on the model parameters. For that, we consider flat priors for each parameter as given in Table \ref{priors}. The priors for $H_0$, $\Omega_{\rm m}$ and $v_{\rm i}$ are guided by \cite{2025Ap&SS.370...62B}, whereas the prior for deviation parameter, $b$, is taken from \cite{kumar2025joint}. Our study is implemented through the emcee\footnote{https://emcee.readthedocs.io/en/stable/} sampler. As number of steps to cover the posterior space are increased from 16000 to 20000, we get converging results. We quote here results obtained by using 20 walkers of 25,000 steps each. In order to avoid biasing due to initial steps and for better converging results, the number of burn-in-steps is taken to be 5\% of the total steps. To track convergence of our MCMC analysis, we further use Gelman-Rubin statistics (denoted by $\hat{R}$) \mbox{(\citealt{gelman1992inference})}. When $\hat{R}$ is close to one, GR diagnostic declares convergence \mbox{(\citealt{2018arXiv181209384V})}. Using the package ArviZ, we test our fits. For all our fits, we find 1.00$<\hat{R}<$ 1.05. We have used GetDist package to plot the marginalized 1D and 2D density plots for the parameters. The GetDist package uses Kernel Density Estimation (KDE) method. The detailed explanation of GetDist package can be found in \cite{2025JCAP...08..025L}. Using this package, one can generate median values, credible intervals and the marginalized densities for the parameters. The scale factors used for 1D and 2D plots are 0.3 and 0.8 respectively.

\begin{table}[H]
\begin{center}
\caption{\small{Priors for the Parameters.}}
    \label{priors}
    \renewcommand{\arraystretch}{1} 
 \begin{tabular}{cc}
      \hline\noalign{\smallskip}
       Parameter  & Prior Range  \\
       \hline\noalign{\smallskip}
       $H_{0}$ & U(50,100)\\ 
       $b$ & U(-1.5,1.5)   \\
       $\Omega_{m}$ & U(0,1)  \\
       v$_{i}$ & U(500,12000)  \\
        \hline\noalign{\smallskip}
\end{tabular}
\end{center}
\end{table}
\section{Results}
\label{Results}
\begin{figure}[ht]
	\centering
    \includegraphics[height=10cm,width=10cm]{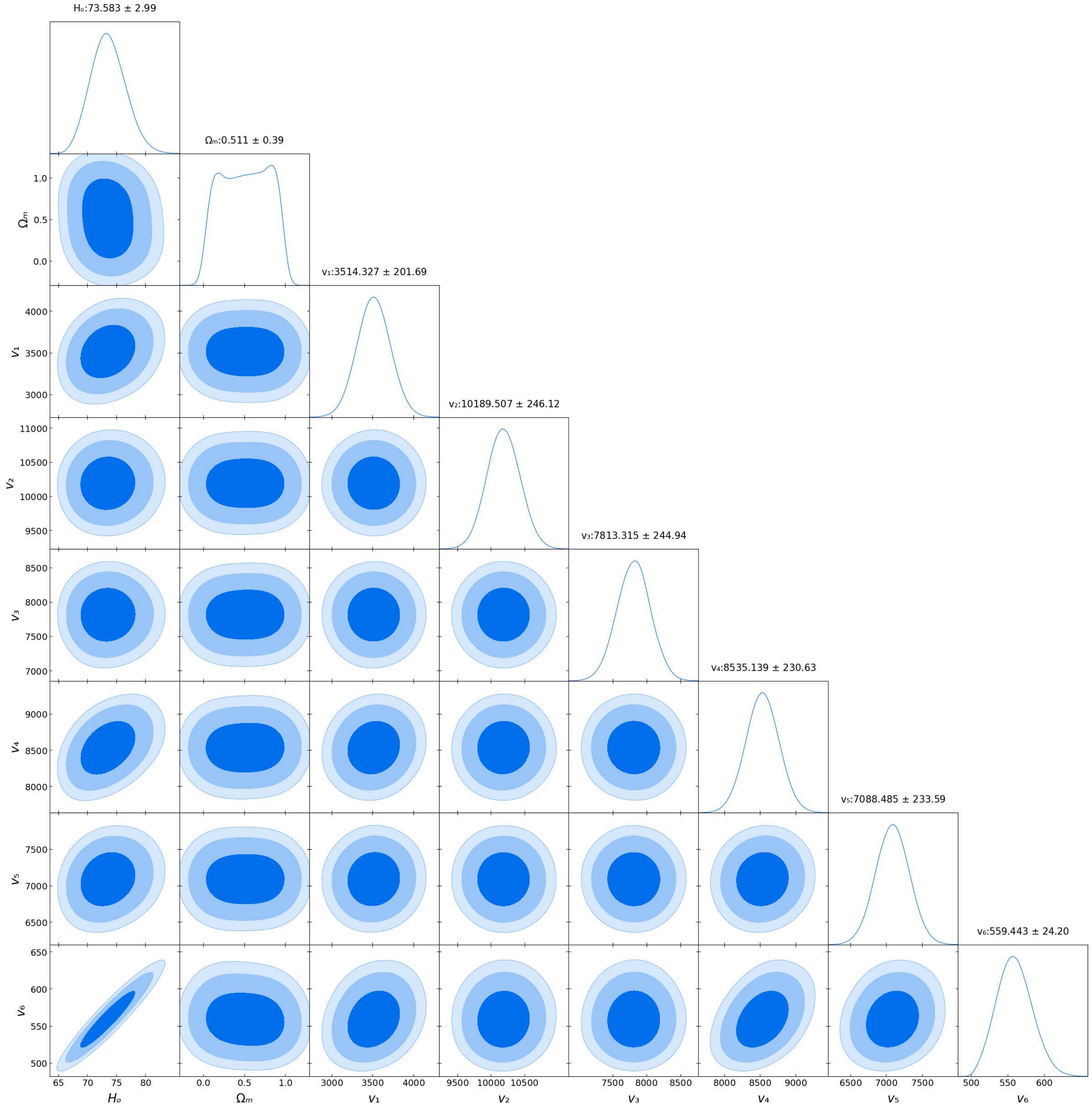}
  	\caption{Constraints on the parameters using emcee for the combined likelihood. Darker to lighter color grades represent 1$\sigma$, 2$\sigma$ and 3$\sigma$ credible contours respectively for the parameters that include $H_{0}$, $\Omega_{\rm m}$ and $v_{i}$ for the  $\Lambda$CDM model. .}
  	\label{CDM model}
\end{figure}

\begin{figure}[ht]
	\centering
    \includegraphics[height=5cm,width=5cm]{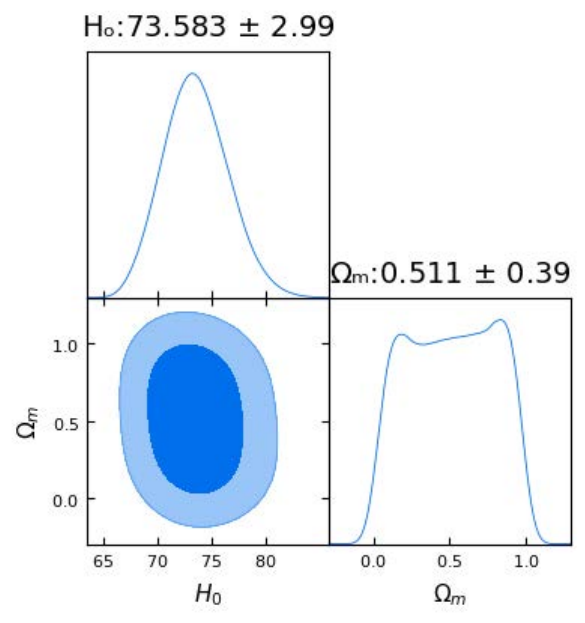}
  	\caption{1$\sigma$ and 2$\sigma$ confidence contours and posterior distributions for the $\Lambda$CDM parameters, i.e., $H_{0}$ and $\Omega_{\rm m}$.}
  	\label{CDM model parameter} 
\end{figure}

The figures present 1D and 2D marginalized plots for the four models, which include the $\Lambda$CDM model along with the three $f(R)$ models. For the $\Lambda$CDM model, we constrain eight parameters ($H_0$, $\Omega_{\rm m}$, $v_1$, $v_2$, $v_3$, $v_4$, $v_5$, $v_6$). Figure \ref{CDM model} shows the confidence contours for the model parameters along with the velocities of the galaxies. Figure \ref{CDM model parameter} is a slice of Figure \ref{CDM model} to draw attention to the model parameters. The marginalized parameter estimates obtained in this work are in concordance with \cite{2025Ap&SS.370...62B}.\\

\begin{figure}[ht]
	\centering
    \includegraphics[height=10cm,width=10cm]{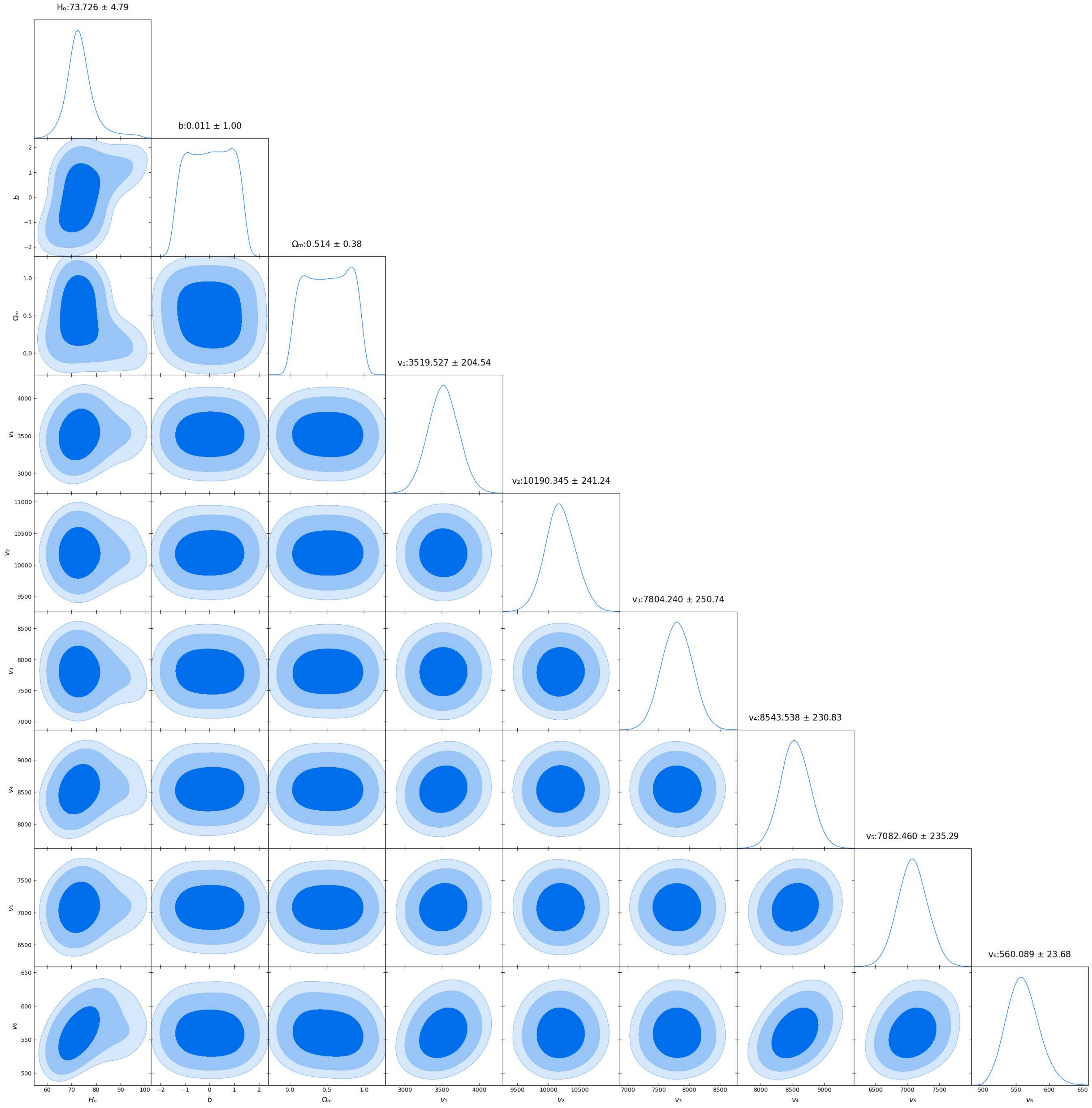}
  	\caption{Constraints on the parameters using emcee for the combined likelihood. Darker to lighter color grades represent 1$\sigma$, 2$\sigma$ and 3$\sigma$ credible contours respectively for parameters, i.e., $H_{0}$, $b$, $\Omega_{\rm m}$ and the six recession velocities in the Hu-Sawicki model.}
  	\label{Hu-Sawicki model}
\end{figure}

\begin{figure}[ht]
	\centering
    \includegraphics[height=10cm,width=10cm]{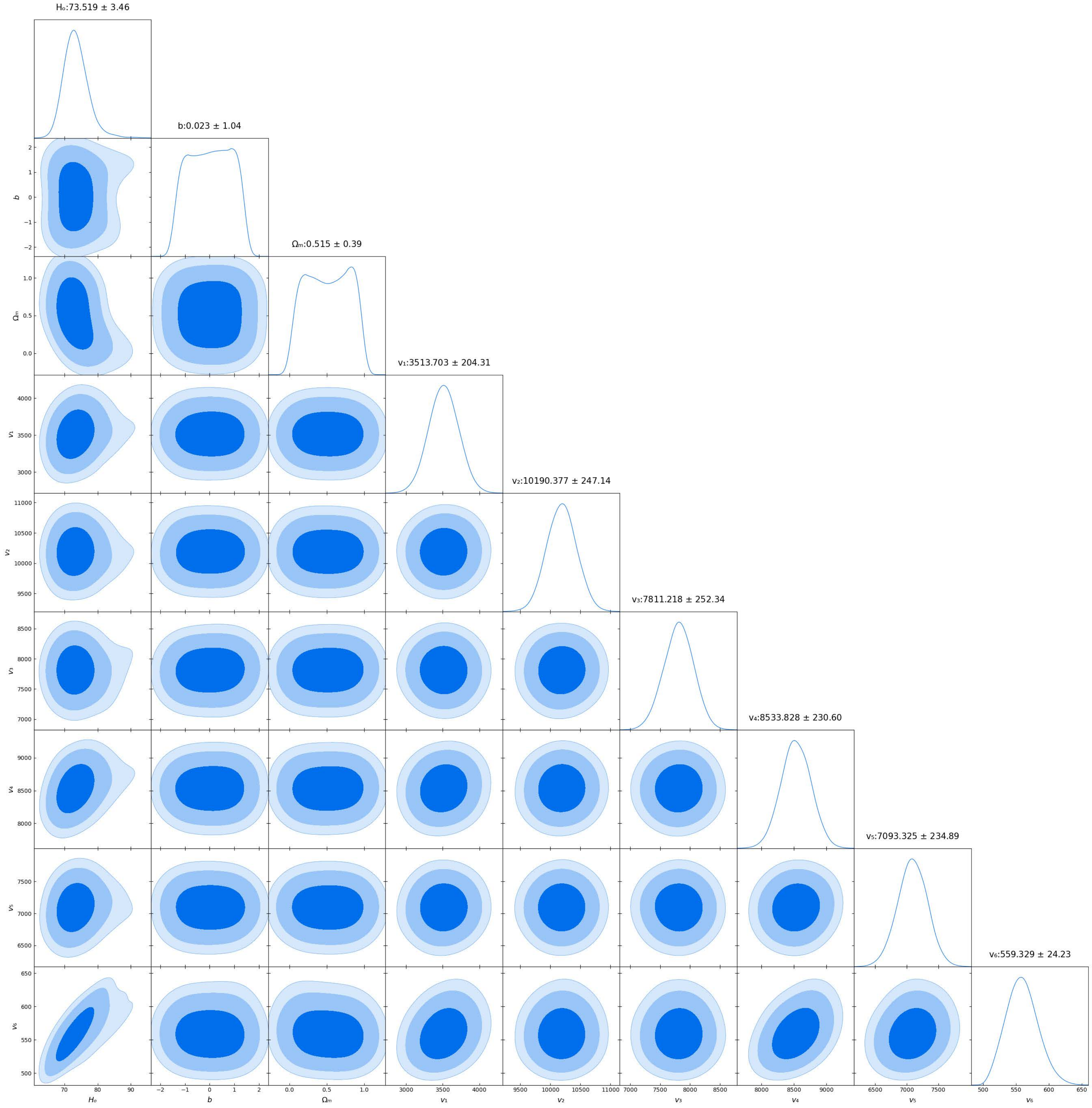}
  	\caption{Constraints on the parameters using emcee for the combined likelihood. Darker to lighter color grades represent 1$\sigma$, 2$\sigma$ and 3$\sigma$ credible contours respectively for the parameters, i.e., $H_{0}$, $b$, $\Omega_{\rm m}$ and the six recession velocities in the Starobinsky model.} 
  	\label{Starboinsky model}
\end{figure}

\begin{figure}[ht]
	\centering
    \includegraphics[height=10cm,width=10cm]{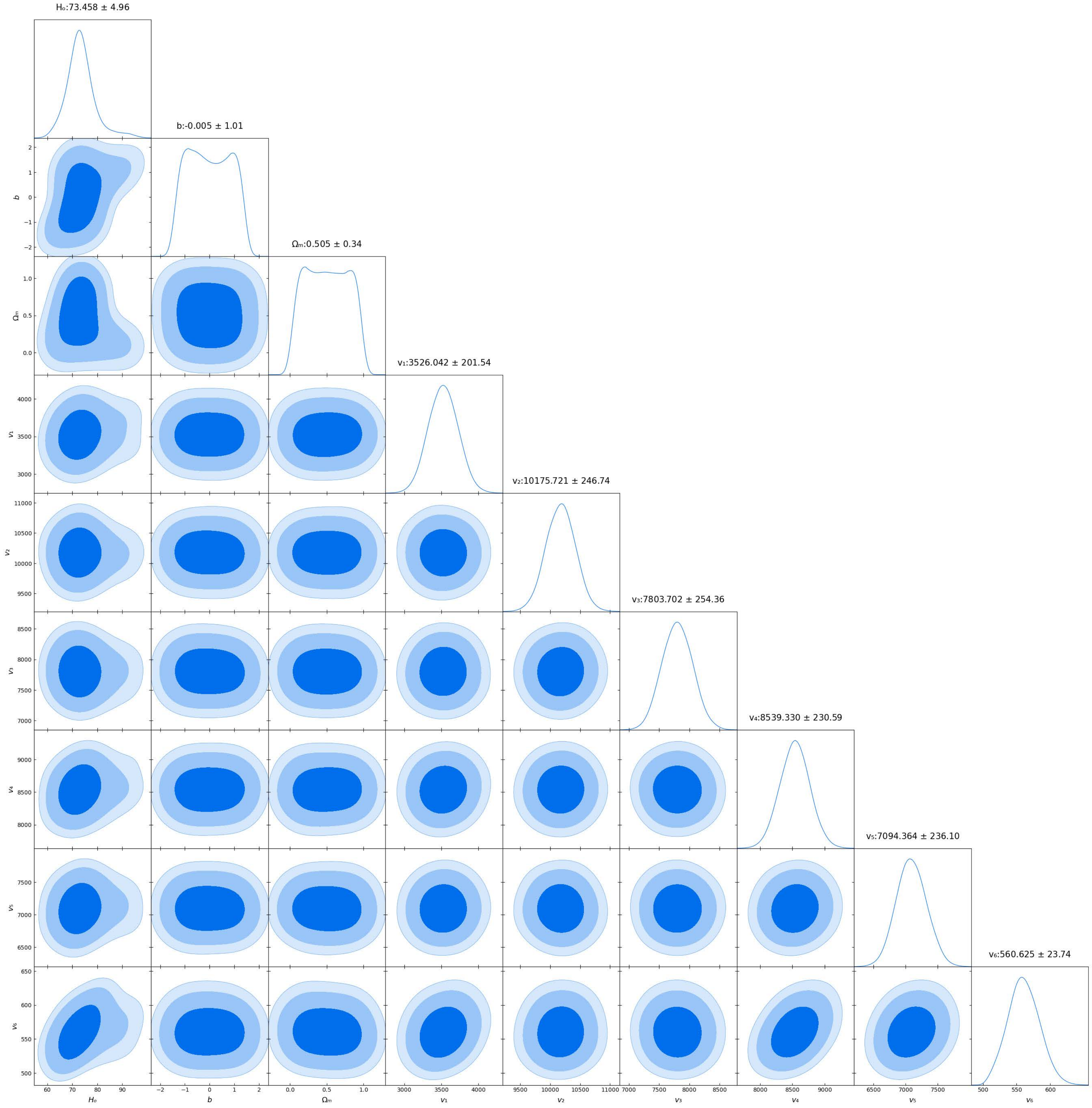}
  	\caption{Constraints on the parameters using emcee for the combined likelihood. Darker to lighter color grades represent 1$\sigma$, 2$\sigma$ and 3$\sigma$ credible contours respectively for the parameters,  i.e., $H_{0}$, $b$, $\Omega_{\rm m}$ and the recession velocities in the ArcTanh model.}
  	\label{ArcTanh model}
\end{figure}

\begin{figure}[ht]
     \centering
     \begin{subfigure}[b]{0.45\textwidth}
         \centering
         \includegraphics[height=4cm,width=4cm]{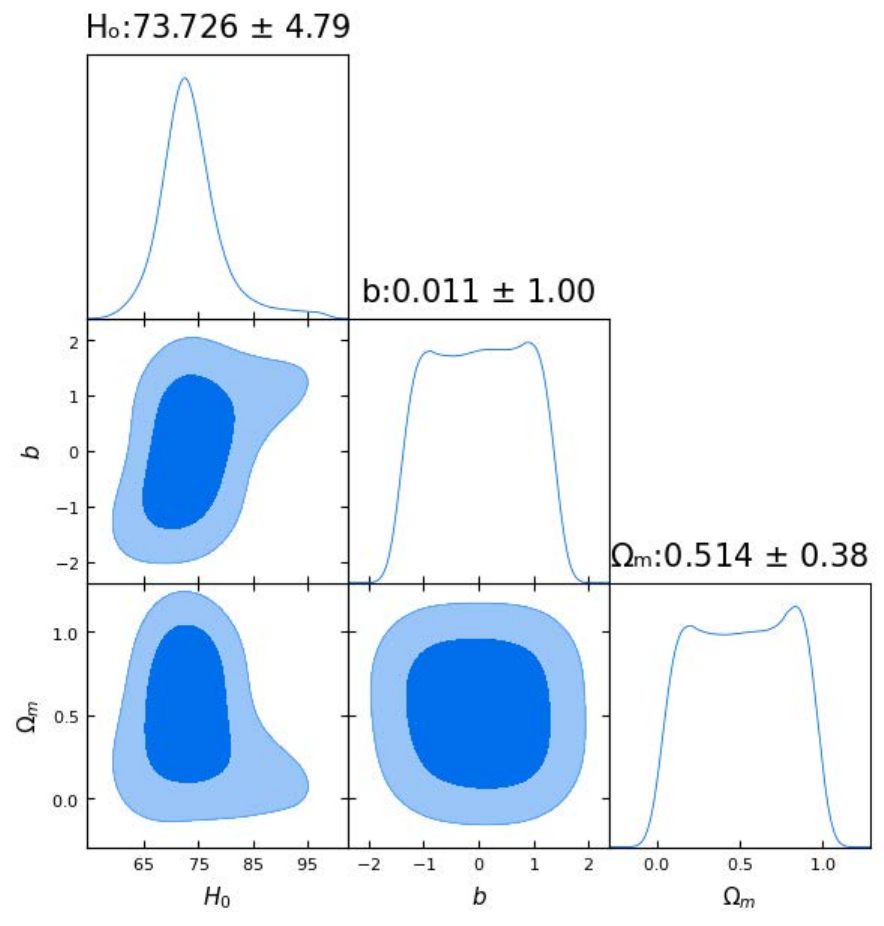}
  	\caption{Hu-Sawicki model}
  	\label{Hu-Sawicki model param}
     \end{subfigure}
     \hfill
     \begin{subfigure}[b]{0.45\textwidth}
         \centering
        \includegraphics[height=4cm,width=4cm]{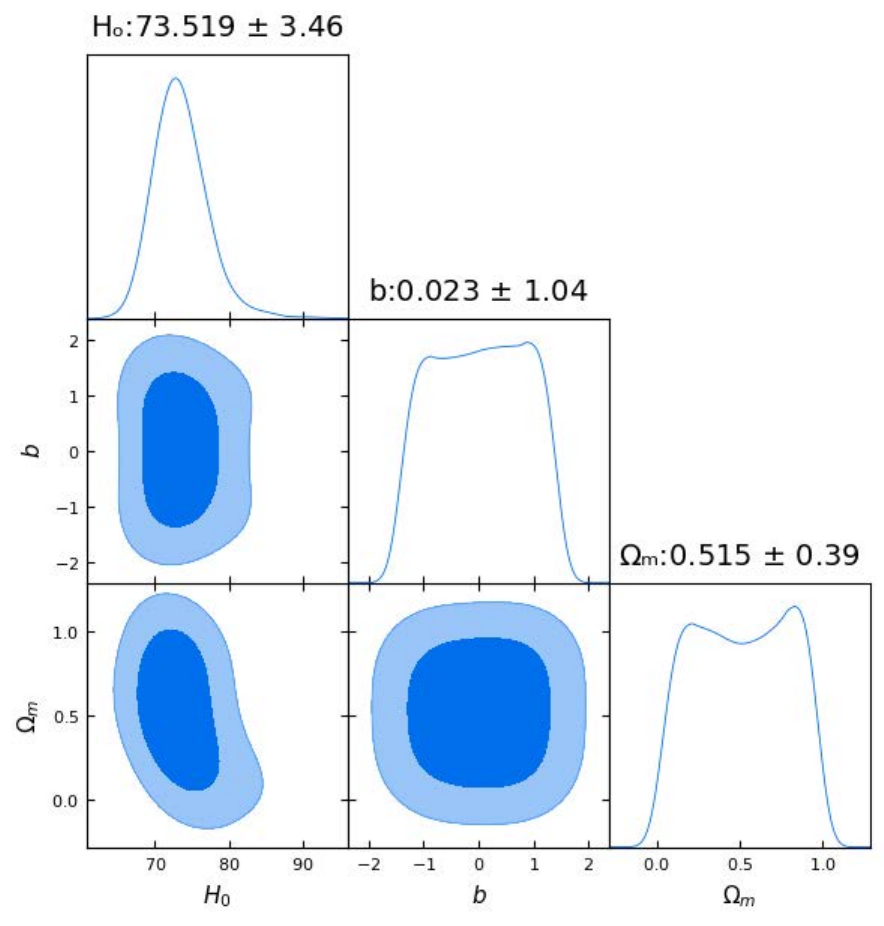}
  	\caption{Starobinsky model}
  	\label{Starobinsky model param}
     \end{subfigure}
      \vspace{10pt}  
     \begin{subfigure}[b]{0.45\textwidth}
         \centering
         \includegraphics[height=4cm,width=4cm]{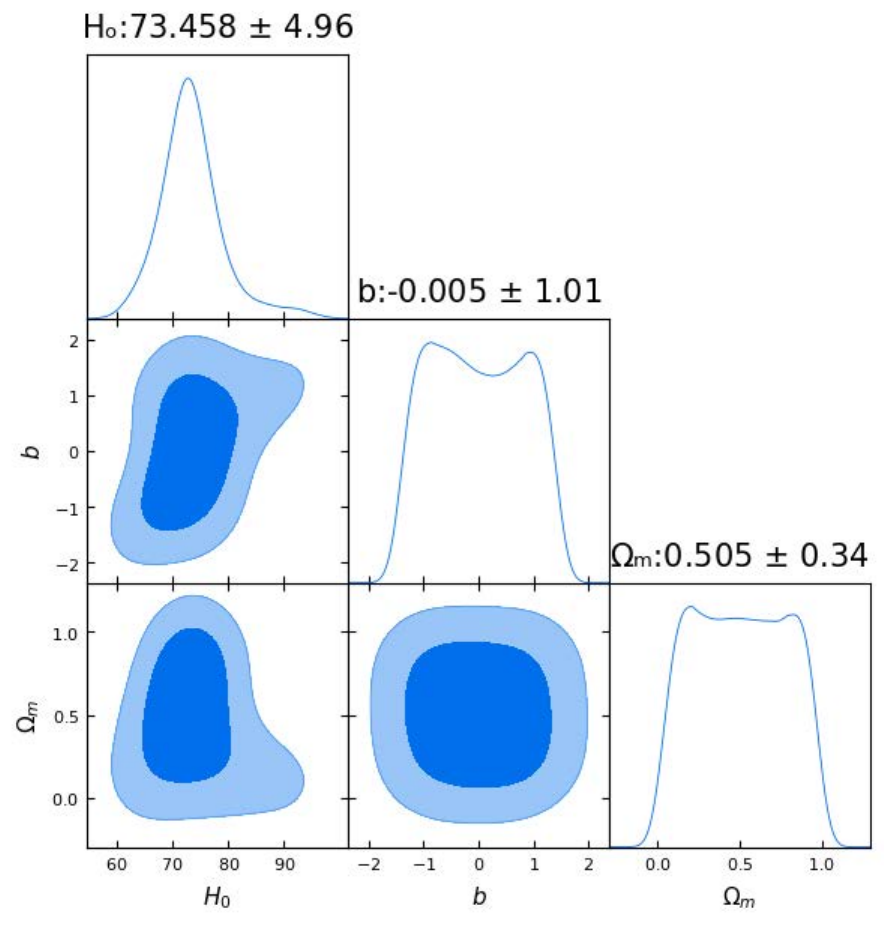}
  	\caption{ArcTanh model}
  	\label{ArcTanh model param}
     \end{subfigure}
     \caption{Marginalized 1$\sigma$ and 2$\sigma$ credible intervals for the model parameters $H_{0}$, $b$ and $\Omega_{\rm m}$ for the three $f(R)$  models. To bring model parameters in focus, we slice Figures \ref{Hu-Sawicki model}, \ref{Starboinsky model} and \ref{ArcTanh model} to produce sub-plots \ref{Hu-Sawicki model param}, \ref{Starobinsky model param} and \ref{ArcTanh model param} respectively.}
     \label{3 parameter fitting}
\end{figure}

\indent Figures \ref{Hu-Sawicki model}, \ref{Starboinsky model} and \ref{ArcTanh model} show the marginalized posterior plots of the parameters, which include $v_i$ and model parameters for the three $f(R)$ gravity models, i.e.,  Hu-Sawicki model, Starobinsky and ArcTanh model respectively. For the three $f(R)$ models, bounds are found on nine parameters ($H_0$, $b$, $\Omega_{\rm m}$, $v_1$, $v_2$, $v_3$, $v_4$, $v_5$, $v_6$) using the  P20 dataset. Figure \ref{3 parameter fitting} represents 1D and 2D density plots for the model parameters $H_{0}$, $\Omega_{\rm m}$ and $b$.

\FloatBarrier
\indent For our study, we have considered the $\sigma_{pec}$ to be 250\,km\,s$^{-1}$. As discussed in Section \ref{Bayesian Analysis}, this is to account for the maximum uncertainty owing to the peculiar motion of galaxies at low redshifts. In order to understand the impact of a fixed value of  $\sigma_{pec}$ being 250\,km\,s$^{-1}$ for all the six host galaxies, we study constraints obtained when, for each galaxy, we assign a   ${\sigma_{pec}}_{i}\, \sim U(150, 250)$. We find that there is not much change in the constraints on the parameters. We check outputs of at least 10 such runs, every time  ${\sigma_{pec}}_{i}\, \sim\, U(150, 250)$. However, as this method is stochastic in nature, it adds a layer of uncertainty which can be combated only by increasing the number of such runs and then working with ensemble average. While the assumption that $\sigma_{pec}$= 250\,km\,s$^{-1}$ for all the host galaxies is a simplification, it not only accounts for the maximum uncertainty owing to peculiar velocity, but is sufficient for comparing cosmological models. We, therefore, continue with this assumption for the study.  \\
\indent Table \ref{results} summarizes the marginalized 1$\sigma$ central estimates of cosmological model parameters  studied in this work. For all the models, constraints on $H_0$ are well-defined and converging. Further the value of $H_0$ is consistent with the value obtained from other low-redshift data (\citealt{Nunes_2017,2022ApJ...938..110B,2023PDU....4201281K,2025,kumar2025joint}). However, $\Omega_{\rm m}$  remains similarly unbounded for all the models.\\
\indent Table \ref{velocities} gives the marginalized 1$\sigma$ central estimates of the recession velocities in the models studied here. We see that the preferred values are converging, i.e., the recession velocity of any given galaxy remain almost same across the models and the observed recession velocity are within 1$\sigma$ or less. \\
\begin{table}[H]
\begin{center}
\caption{\small{The Marginalized Parameter Estimates for the Four Models.}}
             \label{results}
             \renewcommand{\arraystretch}{1} 
 \begin{tabular}{lcc}
 \hline\noalign{\smallskip}
        		Model & Parameters & Marginalized Parameter Estimates     \\
        		\hline\noalign{\smallskip}
        		$\Lambda$CDM  &$H_{0}$   & 73.583 $\pm$ 2.99     \\
        		 
        		&$\varOmega_{m}$ &       0.511 $\pm$ 0.39            \\
        		 
        		Hu-Sawicki & $H_{0}$& 73.726 $\pm$ 4.79 \\
        		 
        		& $\varOmega_{m}$ &0.514 $\pm$ 0.38   \\
        		  
        		& b &0.011 $\pm$ 1.00 \\
        		 
        		Starobinsky & $H_{0}$ &73.519 $\pm$ 3.46   \\
        		 
        		& $\varOmega_{m}$ &0.515 $\pm$ 0.39  \\
        		  
        		& b & 0.023 $\pm$ 1.04 \\
        		 
        		ArcTanh & $H_{0}$ & 73.458 $\pm$ 4.96\\
        		 
        		& $\varOmega_{m}$ &0.505 $\pm$ 0.34  \\
        		 
        		& b &-0.005 $\pm$ 1.01 \\
        		\hline\noalign{\smallskip}
\end{tabular}
\end{center}
\tablecomments{0.86\textwidth}{The units of $H_0$ are km\,s$^{-1}$\,Mpc$^{-1}$.}
\end{table}
\begin{table}[ht]
\begin{center}
 \caption{\small{Bounds  Obtained on Recession Velocities of the Host Galaxies for the Four Models.}}
            \label{velocities}
             \renewcommand{\arraystretch}{1} 
        	\begin{tabular}{ccccc}
        		 \hline\noalign{\smallskip}
        		Velocity & $\Lambda$CDM  & Hu-Sawicki  & Starobinsky &ArcTanh   \\
        		 \hline\noalign{\smallskip}
        		v$_{1}$  &3514.33 $\pm$ 201.69& 3519.53 $\pm$ 204.54&3513.70 $\pm$ 204.31&3526.04 $\pm$ 201.54\\
                v$_{2}$  &10189.51 $\pm$ 246.12& 10190.34 $\pm$ 241.24&10190.38 $\pm$ 247.14&10175.72 $\pm$ 246.74\\
                v$_{3}$  &7813.31 $\pm$ 244.94&7804.24 $\pm$ 250.74&7811.22 $\pm$ 252.34&7803.70 $\pm$ 254.36\\
                v$_{4}$  & 8535.14 $\pm$ 230.63&8543.54 $\pm$ 230.83&8533.83 $\pm$ 230.60&8539.33 $\pm$ 230.59\\
                v$_{5}$  &7088.48 $\pm$ 233.59& 7082.46 $\pm$ 235.29&7093.32 $\pm$ 234.89& 7094.36 $\pm$ 236.10\\
                v$_{6}$  & 559.44 $\pm$ 24.20& 560.09 $\pm$ 23.68&559.33 $\pm$ 24.23&560.62 $\pm$ 23.74\\
        		 \hline\noalign{\smallskip}
        	\end{tabular}   
           \end{center}
 \end{table}
\begin{figure}[ht]
     \centering
     \begin{subfigure}[b]{0.48\textwidth}
         \centering
        \includegraphics[height=5cm,width=6cm]{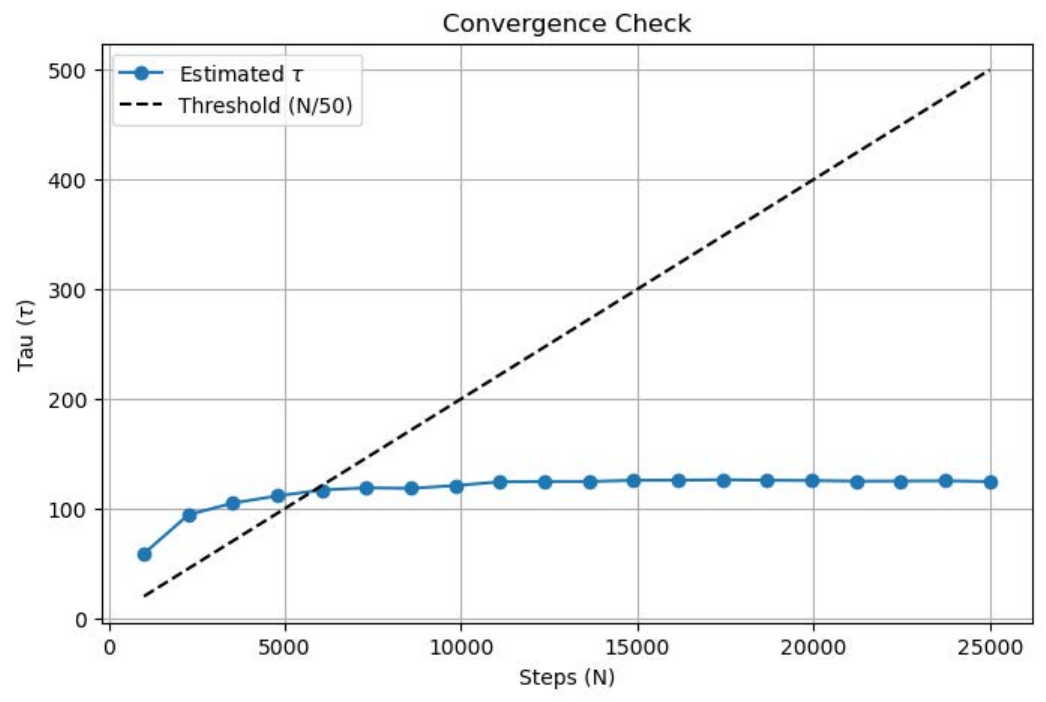}
  	\caption{$\Lambda$CDM model}
     \end{subfigure}
      \vspace{10pt} 
     \begin{subfigure}[b]{0.48\textwidth}
         \centering
         \includegraphics[height=5cm,width=6cm]{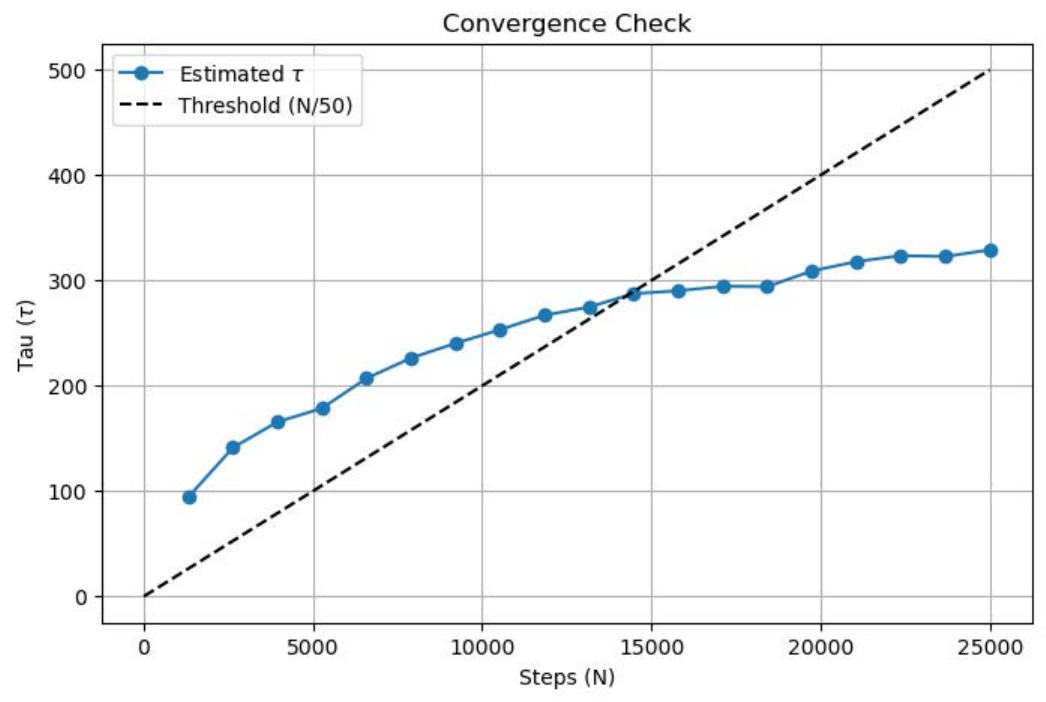}
  	\caption{Hu-Sawicki model}
     \end{subfigure}
     \hfill
     \begin{subfigure}[b]{0.48\textwidth}
         \centering
        \includegraphics[height=5cm,width=6cm]{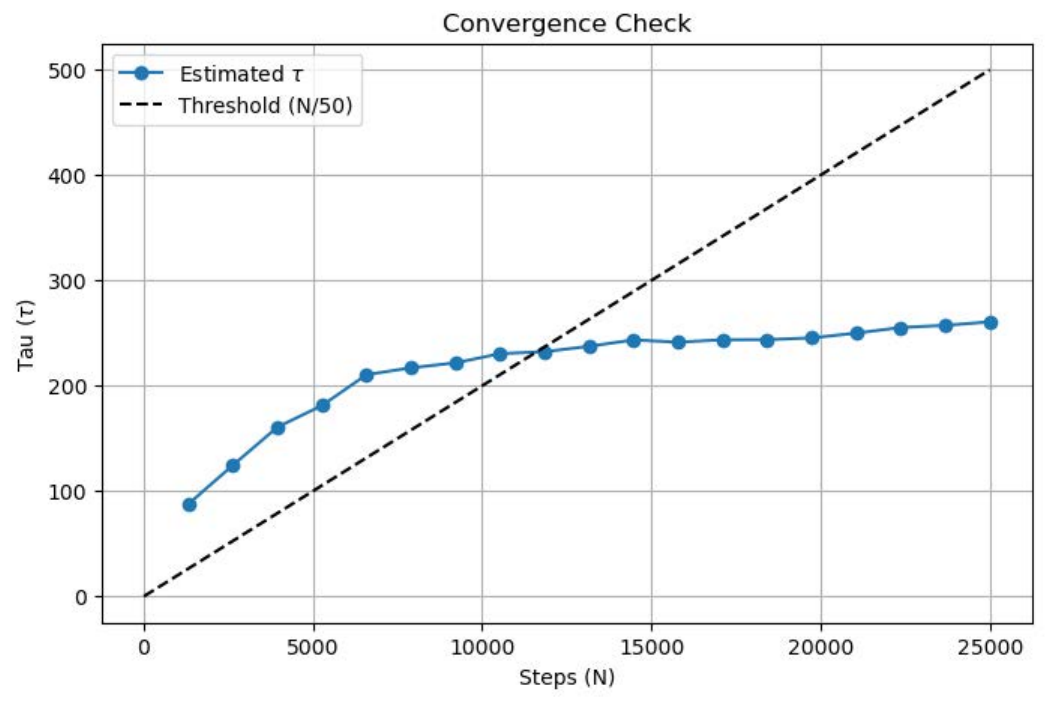}
  	\caption{Starobinsky model}
     \end{subfigure}
      \vspace{10pt}  
     \begin{subfigure}[b]{0.48\textwidth}
         \centering
         \includegraphics[height=5cm,width=6cm]{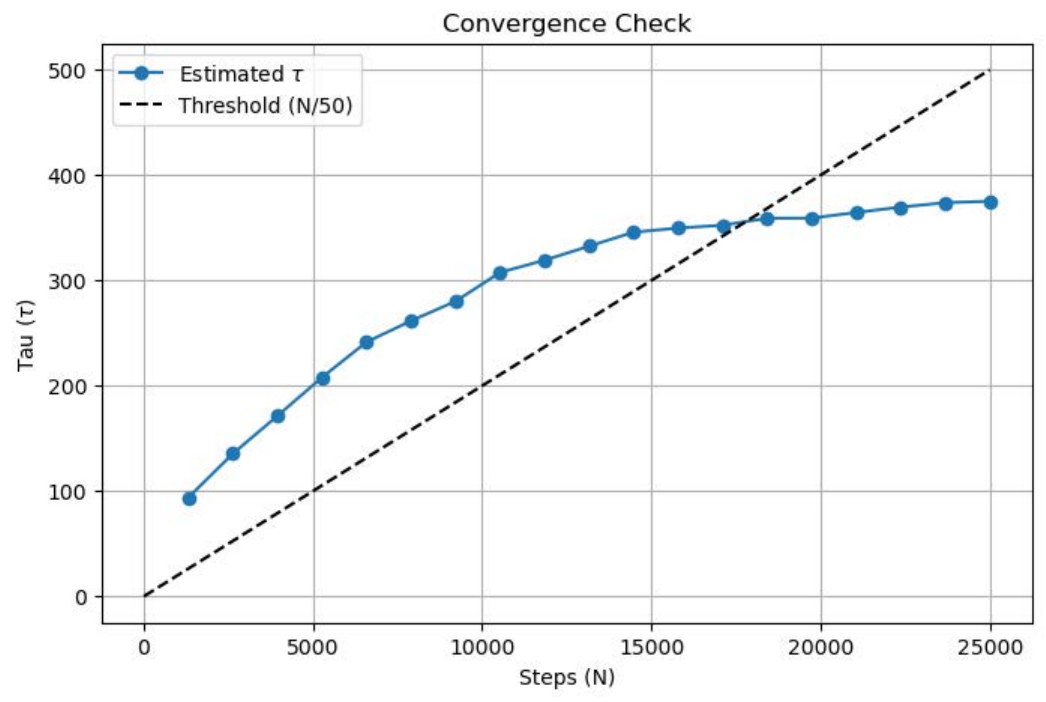}
  	\caption{ArcTanh model}
     \end{subfigure}
     \caption{Plots to verify the convergence of MCMC chains  using auto-correlation time function, $\tau$. The blue line represents the variation of $\tau$ with number of steps (N) to show convergence. The black dashed line represents the threshold (N/50) for the MCMC runs \mbox{(\citealt{2024MNRAS.531.4181O})}. Once the $\tau$ versus N curve crosses the threshold line, it indicates that the chains start to get more reasonable (convergent).}
     \label{convergence plots}
     
\end{figure}
\indent We use the auto-correlation time, $\tau$, to verify our results. This integrated auto-correlation time function of emcee\footnote{https://emcee.readthedocs.io/en/stable/user/autocorr/} is a marker to check independence and convergence of the samples. The results are considered to be efficient if $\tau< N/50$ (where N is the number of steps in the sampled chains) \mbox{(\citealt{2024MNRAS.531.4181O})}. Figure \ref{convergence plots}  shows our convergence plots for the four models. 
\FloatBarrier
 
\subsection{Model Comparison}
\label{Model Comparison}
\noindent To compare the models studied in this work, we employ widely used information criteria, Akaike Information Criterion (AIC) and Bayesian Information Criterion (BIC). The criteria  depend on the value of $\chi^{2}_{\rm min}$, number of free parameters of the model, $d$ and the number of data points, $n$. AIC and BIC are respectively defined as:
\begin{equation}
     {{\rm AIC}=\chi^{2}_{\rm min} +{2d}\quad \text{and}} 
\end{equation}
\begin{equation}
     {\rm BIC}=\chi^{2}_{\rm min} +{d\,\,ln(n)}.
\end{equation}
The model with lower value of AIC or BIC is usually considered to be favored over the other models (\citealt{2022MNRAS.514.5827S}). Further, we compare the $f(R)$ models with respect to the $\Lambda $CDM model. For this, we calculate $\Delta \chi_{\rm min}^{2}$, $\Delta$AIC and $\Delta$ BIC, which are defined as:
\begin{equation}
 \Delta \chi_{\rm min}^{2}=(\chi_{\rm min}^{2})_{\text{$f(R)$ model}} -(\chi_{\rm  min}^{2})_{\Lambda \rm CDM} \quad \text{and} 
\end{equation}
\begin{equation}
 \Delta X=X_{\text{$f(R)$ model}} -X_{\Lambda \rm  CDM}.
\end{equation}

\noindent Here, $X$ stands for AIC$/$BIC.\\
\indent A summary of model comparisons, treating $\Lambda$CDM as a reference model, is presented in Table \ref{AIC_BIC}. As, $0\leq |\Delta X| \leq 2$, it shows that P20 dataset finds $f(R)$ models statistically indistinguishable from $\Lambda$CDM. Also, all the three $f(R)$ models are equally favoured.
 \begin{table}[ht]
\begin{center}
\caption{ For $\Lambda$CDM model, $d$ = 8 Whereas for the $f(R)$ Models, $d$ = 9.}
            \label{AIC_BIC}
            \setlength{\tabcolsep}{3pt} 
            \renewcommand{\arraystretch}{1} 
        	\begin{tabular}{lcccccc}
        		 \hline\noalign{\smallskip}
        		Model    & ${\chi^{2}_{\nu}}$& $\Delta \chi^{2}_{\rm min}$ &AIC & $\Delta$AIC & BIC & $\Delta$BIC    \\
                \hline\noalign{\smallskip}
        		$\Lambda$CDM  &0.65&0    &18.60&0&22.48
        		& 0  \\
        		 
        		Hu-Sawicki  &0.92&0.15&20.75&2.15&25.12&2.64\\
        	     
        	    Starobinsky & 0.92&0.16&20.76&2.16& 25.13&2.65 \\
        	 	 
        	    ArcTanh  &  0.92&0.15&20.76&2.15&25.12& 2.64.\\
        		 \hline\noalign{\smallskip}
        	\end{tabular}   
            \end{center}
\tablecomments{0.7\textwidth}{As we use combined likelihood analysis (see Equation \eqref{chi2}), $n$=12. The second column gives the chi squared per degree of freedom for each model. Here, $\Delta$AIC and $\Delta$BIC test f(R) models against $\Lambda$CDM.}
\end{table} 
\section{Conclusions}
\label{conclusion}
\begin{figure}
	\centering
    \includegraphics[height=9cm,width=13cm]{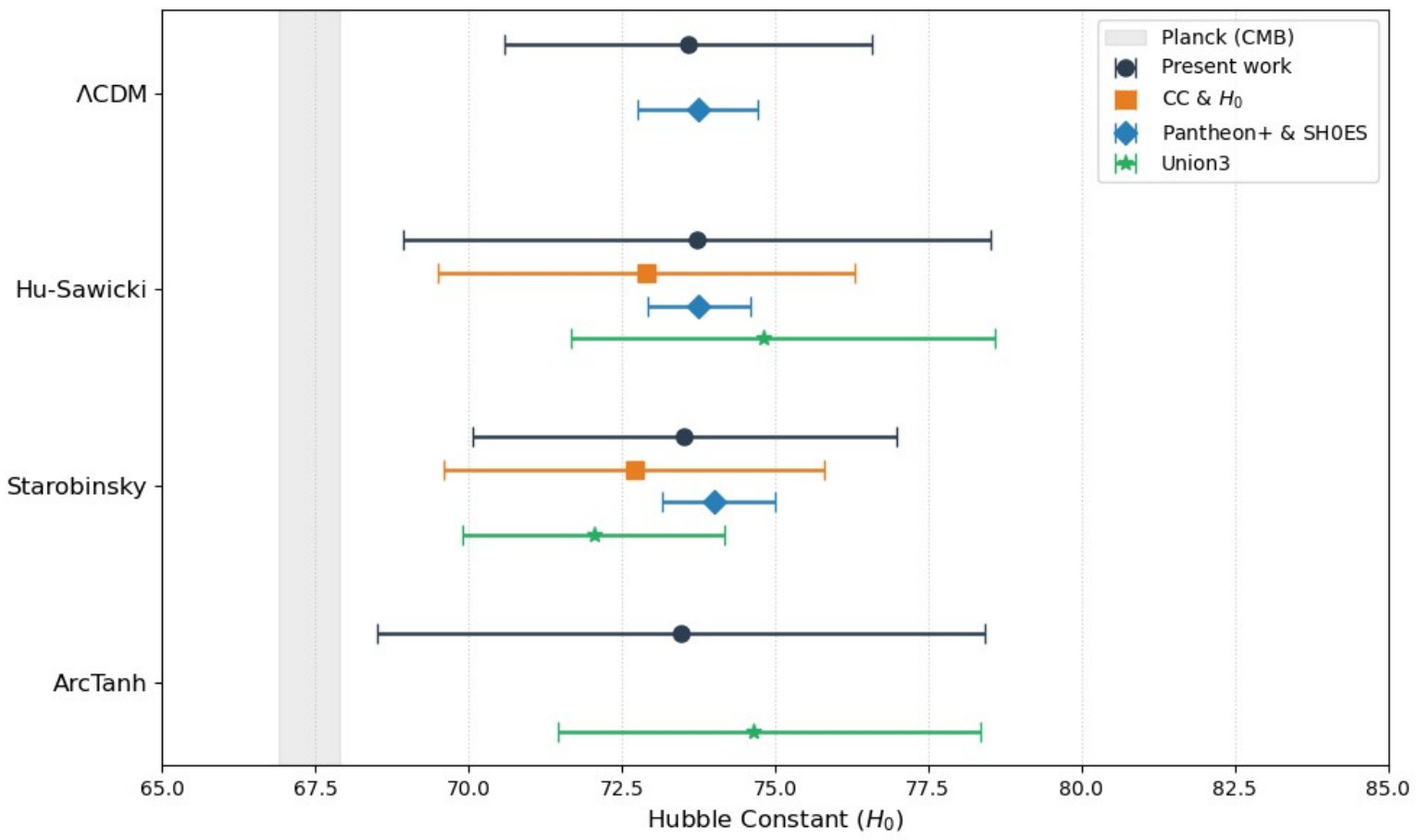}
  	\caption{A comparative study of $H_0$ constraints obtained in this work using the P20 dataset (labeled in black color) with other late-time datasets as mentioned in Table \ref{comparison}. The orange, blue and green colors represent the $H_0$ constraints from the observational datasets CC \& $H_0$, Pantheon+ \& SH0ES and Union3 respectively \mbox{(\citealt{Nunes_2017, 2023PDU....4201281K, kumar2025joint})}. The gray band represents the $H_0$ constraints from Planck observations \mbox{(\citealt{2020A&A...641A...6P})}. }
  	\label{ho_model_comp}
\end{figure}

In this work, we have tested the efficiency of $f(R)$ models using the P20 dataset which is a low redshift data. We constrain recession velocities of the six megamasers along with the model parameters $H_{0}$, $b$ and $\Omega_{\rm m}$ for the three variants of $f(R)$. For reference, we give a summary of the favoured values of the model parameters for the different cosmologies obtained using other late-time datasets in Table \ref{comparison}. References for these results are included in the Table.\\
\indent  We find that $H_0$ for $f(R)$ models provide converging results and the bounds obtained for $H_0$ using P20 dataset are in agreement with those obtained from other datasets (within 1$\sigma$). Figure \ref{ho_model_comp} summarizes the comparison of  the preferred values of $H_0$ obtained in this work with those obtained from (i) the CC+ $H_0$ dataset (\citealt{Nunes_2017}), (ii) the Pantheon+ \& SH0ES dataset (\citealt{2023PDU....4201281K}) and (iii) the Union3 dataset (\citealt{kumar2025joint}). The value of $H_0$ obtained for $\Lambda$CDM model using the early-universe Planck (CMB) (\citealt{2020A&A...641A...6P}) dataset is shown in gray band. This is to highlight the Hubble constant discrepancy.\\
\indent The marginalized estimates of the deviation parameter, $b$, for the three $f(R)$ gravity models are close to zero, even though the constraints obtained are not very restrictive. As shown in Table \ref{comparison} for the three f(R) models, the value of deviation parameter within 1$\sigma$ is consistent with the preferred value of $b$ obtained by \mbox{\cite{Nunes_2017,2023PDU....4201281K,kumar2025joint}}.
 This parameter serves as a quantitative measure of deviation from the standard $\Lambda$CDM cosmology. In the limit $b \longrightarrow0$, these models are constructed to reproduce the $\Lambda$CDM to ensure consistency with current observational constraints \mbox{(\citealt{2007JETPL..86..157S,2010LRR....13....3D})}. On the other hand, in the limit b$\longrightarrow \infty$, the functional form $f(R)$ reduces to $R$. \\
\indent The fact that the inferred values of $b$ remain close to zero indicates that the current dataset does not provide statistically significant evidence for deviation from $\Lambda$CDM. This behavior is expected, as viable $f(R)$ models are typically designed to closely mimic the $\Lambda$CDM background evolution at low redshifts, with deviations primarily manifesting in the growth of cosmic structures rather than in the expansion history \mbox{(\citealt{2010LNP...800...99T})}. Hence, our results indicate that, within the uncertainties of the available data, these modified gravity models remain statistically indistinguishable from $\Lambda$CDM. \\
\indent Results obtained in this paper show that the $\Omega_{\rm m}$ remains poorly constrained for all the models. Besides the fact that the P20 dataset is currently very small, the limitation may be predominately arising from the fact that it is a low-redshift data ($z<<$1). This is corroborated by the constraints obtained on $\Omega_m$ for the three $f(R)$ models and $\Lambda$CDM. Table \ref{comparison} summarizes constraints on $\Omega_m$ for the models under study using CC \& $H_0$, Pantheon+ \& SH0ES and Union3 \mbox{(\citealt{Nunes_2017, 2023PDU....4201281K, kumar2025joint})}. It is evident that the constraints from other datasets/combined datasets are more profound with its conventional value of 0.27 within 1$\sigma$. It has been discussed earlier by \mbox{\cite{1999astro.ph..5116H}} that in the distance measurement, $\Omega_{\rm m}$ enters only at higher-order terms in $z$, making low-redshift data intrinsically insensitive to $\Omega_{\rm m}$. Also at low redshift, peculiar velocities of the galaxies are comparable to Hubble flow. This introduces significant scatter in distance measurements causing weakening of the matter density parameter \mbox{(\citealt{2006PhRvD..73l3526H})}. Constraints on cosmological parameters may become more conclusive by combining other high redshift data like SNe Ia, BAO and H(z).\\
\indent The comparison of the three models using the information criteria, AIC and BIC, show that the P20 dataset finds $f(R)$ models statistically equivalent to $\Lambda$CDM. This is in concordance with the fact that marginalized estimate for the deviation parameter $b$ remains close to zero for all three $f(R)$ models (see Table \ref{results}).\\
\indent Use of megamaser data seems to pose a limitation as the number of parameters (galaxy velocities and model parameters) forms rather huge domain. It is constrained by the fact that uncertainties owing to peculiar velocities have to be addressed for low-redshift distance indicators. \mbox{\cite{2020ApJ...891L...1P}} develop various methods of addressing uncertainties due to peculiar velocities including the one adopted for our analysis. \mbox{\cite{2020ApJ...891L...1P}} use "leave-one-out" jackknife tests to establish robustness of these treatments. Given the robustness established by the "leave-one-out" test, it becomes important tool for constraining $H_0$ and other cosmological parameters.\\
\indent We believe that the increase in the volume of the megamaser data from the upcoming and ongoing surveys like High-Frequency VLBI and Australian SKA Pathfinder may help us to constrain the model parameters in a more decisive manner (\citealt{2012SPIE.8444E..2AS,Chen_2020}). 
\FloatBarrier
 \begin{table}[h]
 \begin{center}
             \caption{\small{Preferred Values of Model Parameters from Other Late-time Observational Datasets.}}
            \label{comparison}
            \renewcommand{\arraystretch}{1}
        	\begin{tabular}{lccccc}
        		 \hline\noalign{\smallskip}
        		 Model & Parameters & P20 dataset$^a$ & \begin{tabular}[c]{@{}c@{}}CC \& $H_0$ [1]\end{tabular} & \begin{tabular}[c]{@{}c@{}}Pantheon+ \& SH0ES [2]\end{tabular} & Union3 [3] \\
             \hline\noalign{\smallskip}
                
        		\multirow{2}{*}{$\Lambda$CDM } &$H_0$  & 73.583 $\pm$ 2.99     
                 &---& 73.74 $\pm$ 0.98 & ---      \\
                 
                  &$\Omega_{m}$  & 0.511 $\pm$ 0.39            
                &--- & 0.333 $\pm$ 0.018&--- \\
        		 \hline\noalign{\smallskip}
        		\multirow{3}{*}{Hu-Sawicki } &$H_{0}$  & 73.726 $\pm$ 4.79
                 & 72.9$^{+3.4}_{-3.4}$ &73.76 $\pm$ 0.83 & 74.823$^{+3.777}_{-3.159}$     \\
                 
                  &$\Omega_{m}$  &0.514 $\pm$ 0.38   
                &0.264$^{+0.069}_{-0.058}$ & 0.334 $\pm$ 0.017 & 0.293$^{+0.067}_{-0.061}$\\
                 
                 &$b$  & 0.011 $\pm$ 1.00 & 0.107$^{+0.316}_{-0.158}$ &$-$0.003 $\pm$ 0.029  & 0.473$^{+0.324}_{-0.478}$\\
                 \hline\noalign{\smallskip}
                 \multirow{3}{*}{Starobinsky} &$H_{0}$  & 73.519 $\pm$ 3.46   
                 & 72.7$^{+3.1}_{-3.1}$ &74$^{+1.0}_{-0.85}$ &  72.043$^{+2.128}_{-2.148}$     \\
                 
                  &$\Omega_{m}$  &0.515 $\pm$ 0.39  
               & 0.261$^{+0.065}_{-0.055}$ & 0.332 $\pm$ 0.019 &0.333$^{+0.028}_{-0.031}$ \\
                
                 &$b$  &0.023 $\pm$ 1.04 
                 & 0.229$^{+0.254}_{-0.710}$ &$-$0.037$^{+0.051}_{-0.044}$ &0.935$^{+0.195}_{-0.533}$ \\
                  \hline\noalign{\smallskip}
                 \multirow{3}{*}{ArcTanh} &$H_{0}$  & 73.458 $\pm$ 4.96
               & ---&--- &74.635$^{+3.726}_{-3.170}$       \\
                 
                  &$\Omega_{m}$  & 0.505 $\pm$ 0.34  
               &--- &--- & 0.296$^{+0.074}_{-0.065}$\\
                
                 &$b$  &  -0.005 $\pm$ 1.01 &--- & ---& 0.442$^{+0.326}_{-0.525}$\\
                  \hline\noalign{\smallskip}
        	\end{tabular}
            \tablecomments{0.85\textwidth}{$^{a}$ Present Work\\
            References: [1] \citealt{Nunes_2017}; [2]  \citealt{2023PDU....4201281K}; [3]  \citealt{kumar2025joint}.}
            \end{center}
 \end{table} 

\FloatBarrier
\section*{Acknowledgements}
\label{acknowledgements}
The authors thank the anonymous referee for valuable suggestions and comments which have helped to strengthen the manuscript. N. Rani would like to thank IUCAA Pune, India for providing her Visiting Associateship under which a part of this work was carried out. We thank Darshan Kumar, Akshay Rana and Daulti Rani for useful discussions.   
\bibliographystyle{raa}
\bibliography{main}
 
\end{document}